\documentclass[aps,prd,showpacs,preprintnumbers,nofootinbib,twocolumn]{revtex4}
\usepackage{bm}
\usepackage{latexsym}
\usepackage{dcolumn}
\usepackage{amsmath,amsfonts,amssymb}
\usepackage{graphicx,epsfig}
\usepackage{psfrag}
\usepackage{amsthm}

\def\be {\begin{equation}}
\def\ee {\end{equation}}
\def\bea {\begin{eqnarray}}
\def\eea {\end{eqnarray}}
\def\bc {\begin{center}}
\def\ec {\end{center}}
\def\bfg {\begin{figure}}
\def\efg {\end{figure}}
\def\bi {\begin{itemize}}
\def\ei {\end{itemize}}
\def\nn {\nonumber}

\def\la {\label}
\def\le {\left}
\def\ri {\right}
\def\pa {\partial}
\def\fr {\frac}

\def\no {\noindent}

\def\hs {\hspace}

\def\b  {\beta}

\def\d  {\delta}

\def\beq{\begin{equation}}
\def\eeq{\end{equation}}
\def\br{\begin{eqnarray}}
\def\er{\end{eqnarray}}
\newcommand{\eel}[1] {\label{#1}\end{equation}}

\newcommand{\bdm}{\begin{displaymath}}
\newcommand{\edm}{\end{displaymath}}

\begin{document}

\title{Phenomenological Implications of the Generalized Uncertainty Principle}

\author{Saurya Das$^1$}
\email[email: ]{saurya.das@uleth.ca}

\author{Elias C. Vagenas$^2$}
\email[email: ]{evagenas@academyofathens.gr}

\affiliation{$^1$~Dept. of Physics, University of Lethbridge, 4401
University Drive, Lethbridge, Alberta, Canada T1K 3M4\\
$^2$~Research Center for Astronomy and Applied Mathematics,
Academy of Athens, Soranou Efessiou 4, 11527, Athens, Greece}

\begin{abstract}
Various theories of Quantum Gravity argue that
near the Planck scale, the Heisenberg Uncertainty Principle should
be replaced by the so called Generalized Uncertainty Principle (GUP).
We show that the GUP gives rise to two additional terms in any
quantum mechanical Hamiltonian,
proportional to $\beta p^4$ and $\beta^2 p^6$ respectively, where
$\beta \sim 1/(M_{Pl}c)^2$ is the GUP parameter.
These terms become important at or above the
Planck energy. Considering only the first of
these, and treating it as a perturbation, we show that the GUP affects
the Lamb shift, Landau levels, reflection and transmission coefficients of a
potential step and potential barrier, and the
current in a Scanning Tunnel Microscope (STM).
Although these are too small
to be measurable at present, we speculate on the possibility of
extracting measurable predictions in the future.

\end{abstract}

\maketitle


\section{Introduction} \label{formalism}

Although, there are various approaches to Quantum Gravity, e.g.
String Theory and Canonical Quantum Gravity, to our knowledge,
none of them has made a single prediction which can be experimentally tested at
present (or in the near future). Even if Supersymmetry is observed
in the Large Hadron Collider (LHC), it would at best confirm the
existence of an essential ingredient of the String Theory, and
would hardly be an evidence in favor of the theory itself. Given
this situation, it is important to try to extract testable
predictions. There has been recent attempts in this direction, and
although some of them do compute Quantum Gravity effects, the
smallness of the Planck length (and largeness of the Planck
energy) too often renders these effects minuscule \cite{qgpheno}.
In this paper, we explore a few well understood low energy systems
and show that Quantum Gravity does predict corrections for them.
These corrections are once again, generically quite small to be
measurable. However, we argue that (i) they could signal a new
intermediate length scale between the electroweak and the Planck
scale, and (ii) study of other related systems could give rise to
predictions which can perhaps be tested
\footnote{Some preliminary results in this direction were published in
\cite{sdecvprl}.}.

Our main ingredient is the so-called Generalized
Uncertainty Principle (GUP), which has been argued from various approaches to Quantum
Gravity and Black Hole Physics, using a combination of thought experiments and
series of arguments \cite{guppapers}. These indicate that there exists
a minimum measurable length \cite{Harbach:2003qz}, the Planck length,
$\ell_{Pl} \approx 10^{-33}~cm$.
The prediction is largely model independent, and can be understood
as follows: the Heisenberg Uncertainty Principle (HUP), whereby
uncertainty in position decreases with increasing energies
($\Delta x \sim \hbar/\Delta p$), breaks down for energies close
to the Planck scale, at which point the corresponding
Schwarzschild radius becomes comparable to the Compton wavelength (both
being approximately equal to the Planck length). Higher energies
result in a further increase of the Schwarzschild radius,
resulting in $\Delta x \approx \ell_{Pl}^2\Delta p/\hbar$.
Consistent with the above, the following form of GUP
has been proposed, postulated to hold at all scales
\cite{guppapers}
\bea \Delta x_i \Delta p_i &\geq& \fr{\hbar}{2} [ 1 + \beta
\le((\Delta p)^2 + <p>^2 \ri) \nn \\
&+& 2\beta \le( \Delta p_i^2 + <p_i>^2\ri) ]~,~i=1,2,3 \la{uncert1} \eea
where $[\beta] =(momentum)^{-2}$ and we will assume that
$\beta=\beta_0/(M_{Pl}c)^2= \ell_{Pl}^2/2\hbar^2$ while $M_{Pl}$ is the Planck mass,
$M_{Pl} c^2=$ (Planck energy) $\approx 10^{19}~GeV$. It
is evident that the parameter $\beta_0$ is dimensionless. But,
what determines its value? It is normally assumed that $\beta_0$
is not far from unity. We will see in this article, that on the
one hand, $\beta_0 \approx 1$ renders the effects of Quantum
Gravity on everyday quantum phenomena too small to be measurable.
On the other hand, if one does not impose the above condition {\it
a priori}, current experiments predict large upper bounds on it,
which are consistent with current observations, and may indeed
signal the existence of a new length scale. Note that any new such
intermediate length scale, $\ell_{inter} \equiv \sqrt{\beta_0}
\ell_{Pl}$ cannot exceed the electroweak length scale $\sim
10^{17}~\ell_{Pl}$ (as otherwise it would have been observed),
this tells us that $\b_0$ cannot exceed about $10^{34}$. (The
factor of $2$ in the last term in Eq.(\ref{uncert1}) follows from
Eq.(\ref{com1}) below).

It was shown in \cite{kmm}, using standard methods, that the above
inequality follows from the modified Heisenberg algebra
\be [x_i,p_j] = i \hbar ( \delta_{ij} + \beta \delta_{ij} p^2 +
2\beta p_i p_j )~. \la{com1} \ee
This form ensures, via the Jacobi identity, that
$[x_i,x_j]=0=[p_i,p_j]$ \cite{kempf}.
Note that the above algebra does not admit of a simple representation in
position space. However, defining
\bea x_i &=& x_{0i} \\
p_i &=& p_{0i} \le( 1 + {\beta} p_0^2 \ri) \la{mom1} \eea
where $p_0^2 = \sum\limits_{j=1}^{3}p_{0j}p_{0j}$ and with $x_{0i}, p_{0j}$
satisfying the canonical commutation relations
\be [x_{0i}, p_{0j}] = i \hbar~\delta_{ij},\ee
it is easy to show that Eq.(\ref{com1}) is satisfied, to order $\beta$.
Henceforth, we neglect terms of order $\beta^2$ and higher.

Here,
$p_{0i}$ can be interpreted as the momentum at low energies
(having the usual representation in position space, i.e. $p_{0i} = -i
 \hbar d/d{x_{0i}}$), and $p_{i}$ as that at higher energies.

Using (\ref{mom1}), any Hamiltonian of the form
\bea H &=& \fr{p^2}{2m} + V(\vec r)~,~~ \vec r =(x_1,x_3,x_3) \eea
can be written as \cite{brau}
\bea
H&=& \fr{p_0^2}{2m} + V(\vec r) + \fr{\beta}{m} p_0^4 +
{\cal O}(\beta^2) \la{gupham1} \\
&\equiv& H_0 + H_1 + {\cal O}( \beta^2)    ~, \eea
where
\bea H_0 = \fr{p_0^2}{2m} + V(\vec r) 
~~\mbox{and}~~ H_1 = {\fr{\beta}{m} p_0^4}
= \fr{\beta \hbar^4}{m} \nabla^4
~\la{H0},\eea
where in the last step, we have specialized to the position
representation.
Thus, we see that {\it any} system with a
well defined quantum (or even classical) Hamiltonian $H_0$, is perturbed
by $H_1$, defined above, near the Planck scale. In other words,
Quantum Gravity effects are in some sense universal! It remains to
compute the corrections to various phenomena due to the
Hamiltonian $H_1$. Before we do that, we note that
using (\ref{gupham1}) to write the time-dependent Schr\"odinger equation
\be
H \psi (\vec r, t) = i\hbar \frac{\partial \psi}{\partial t}~,
\la{timedepSE1}
\ee
and going through the usual set of steps (multiplying (\ref{timedepSE1}) by $\psi^\star$,
and subtracting it from the complex conjugate of (\ref{timedepSE1})
multiplied by $\psi$), and making a few further manipulations, one arrives at the following
charge and current densities and the conservation equation
\bea
\vec \nabla \cdot \vec J &+& \frac{\pa \rho}{\pa t} = 0 \\
\rho &=& |\psi|^2 \\
\vec J &=& \frac{\hbar}{2mi} \le[ \psi^\star \vec\nabla\psi - \psi\vec\nabla\psi^\star\ri]
\nn \\
&-& \frac{\beta\hbar^3}{mi}
\le[
\le(\psi^\star \vec\nabla \nabla^2 \psi - \psi \vec\nabla \nabla^2\psi^\star \ri) \ri.  \nn \\
&& \le. +\le( \nabla^2 \psi^\star \vec\nabla \psi - \nabla^2\psi \vec\nabla \psi^\star \ri)
\ri] \la{newcurrent1} \\
&\equiv& \vec{J}_0 + \vec{J}_1~, \la{newcurrent2}
\eea
where $\vec{J}_0$ is the usual quantum mechanical expression and $\vec{J}_1$ is the
additional $\b$-dependent term due to GUP. It is satisfying that the modified
Hamiltonian (\ref{gupham1}) does admit of a (new) conserved current.
Next, we study its effect on a number of quantum mechanical systems,
with various $V(\vec r)$.

\section{GUP and the Lamb shift}

For the Hydrogen atom, $V(\vec r) = -k/r$ ($k=e^2/4\pi
\epsilon_0$, $e=$ electronic charge, $r=\sqrt{|\vec r|}$), for which,
to first order, the perturbing Hamiltonian
$H_1$ shifts the wave-functions to \cite{bransden}
\bea |\psi_{nlm} \rangle_1 = |\psi_{nlm} \rangle +\hspace{-2ex}
\sum_{\{n'l'm'\} \neq \{ nlm\}}\hspace{-1.5ex}
\frac{e_{n'l'm'|nlm}}{E_{n} - E_{n'}} |n'l'm'\rangle \la{wavefn1}
\eea
where $n,l,m$ have their usual significance, and
\be e_{n'l'm'|nlm} \equiv \langle n'l'm'|H_1|nlm\rangle ~.\ee
Using $p_0^2=2m[H_0 + k/r]$, we get \cite{brau}
\be H_1 = (4\beta m) \le[ H_0^2 + k \le( \frac{1}{r}H_0 + H_0
\frac{1}{r} \ri) + \le( \frac{k}{r}\ri)^2 \ri] ~.\ee
Thus,
\bea \frac{ e_{n'l'm'|nlm}}{4\beta m} &=& \le( E_n \ri)^{2}\delta _{n n'}+ k \le( E_n + E_{n'} \ri)
\langle n'l'm'|\frac{1}{r} | nlm \rangle \nn\\
&+& k^2 \langle n'l'm'|\frac{1}{r^2} | nlm \rangle~. \la{mat0}
\eea
From the orthogonality of spherical harmonics, it follows that the
above are non-vanishing if and only if $l'=l$ and $m'=m$. Thus,
the first order shift in the ground state wave-function is given
by (in the position representation)
\bea \Delta \psi_{100}(\vec r) &\equiv& \psi_{100(1)}(\vec
r)-\psi_{100}(\vec r) =
\frac{e_{200|100}}{E_1-E_2}\psi_{200} (\vec r)\nn\\
&=& \frac{928\sqrt{2} \beta m E_0}{81}~\psi_{200}(\vec r) ~, \eea
where we have used the following:\\
(i) the first term in the sum in Eq.(\ref{wavefn1}) ($n'=2$)
dominates, since $E_n = -E_0/n^2~(~E_0=e^2/8\pi \epsilon_0 a_0 =
k/2a_0= 13.6~eV~,~a_0= {4\pi \epsilon_0 \hbar^2}/{me^2} = 5.3
\times
10^{-11}~\mbox{metre}$~,~$m=$ electron mass $=0.5~MeV/c^2$), \\
(ii) $\psi_{nlm} (\vec r) = R_{nl} (r) Y_{lm}(\theta,\phi)$, \\
(iii) $R_{10}=2 a_0^{-3/2}e^{-r/a_0}~
~\mbox{and}~~\\
~~~~~~~
R_{20}=(2a_0)^{-3/2}\le(2 - r/a_0\ri) e^{-r/2a_0}$,\\
(iv) $Y_{00}(\theta,\phi)=1/(\sqrt{4\pi})$,\\
where $E_0$ is the lowest (ground state) energy level of the Hydrogen atom and $a_0$ is the Bohr radius.
\par\noindent
Next, consider the expression for the Lamb shift for the $n^{th}$
level of the Hydrogen atom ($\alpha \equiv e^2/4\pi\epsilon_0\hbar
c \approx 1/137$) \cite{bd}
\be
\Delta E_n = \frac{4\alpha^2}{3m^2} \le( \ln \frac{1}{\alpha} \ri) \le|
\psi_{nlm}(0) \ri|^2~.
\ee
Varying $\psi_{nlm}(0)$, the additional contribution to the Lamb
shift due to GUP in proportion to its original value is given by
\be
\frac{\Delta E_{n(GUP)}}{\Delta E_n} =
2 \frac{\Delta|\psi_{nlm}(0)|}
 {\psi_{nlm}(0)}~.
\ee
Thus, for the Ground State, using $\psi_{100}(0) =
a_0^{-3/2}\pi^{-1/2}$ and $\psi_{200}(0)=
a_0^{-3/2}(8\pi)^{-1/2}$, we get
\bea \frac{\Delta E_{0(GUP)}}{\Delta E_0} &=& 2
\frac{\Delta|\psi_{100}(0)|}
 {\psi_{100}(0)}= \frac{928 \beta m E_0}{81} \nn\\
 &\approx& 10 \b_0 \frac{m}{M_{Pl}}
 \frac{E_0}{M_{Pl} c^2} \nn\\
&\approx& 10 \times \le( 0.42 \times 10^{-22} \ri) \times \le( 1.13 \times 10^{-27}\ri) \b_0\nn\\
 &\approx& 0.47 \times 10^{-48}~\b_0~.
\eea
The above result may be interpreted in two ways. First, if one assumes
$\beta_0 \sim 1$, then it predicts a non-zero, but virtually
{\it unmeasurable} effect of Quantum Gravity/GUP. On the other hand, if such
an assumption is not made, the current accuracy of
precision measurement of Lamb shift of about
$1$ part in $10^{12}$ \cite{brau,newton}, sets the following upper bound on $\b_0$
\be
\b_0 < 10^{36}~.
\la{beta1}
\ee
This bound is weaker than that set by the electroweak scale, but not
incompatible with it.
Moreover, with more accurate measurements in the future, this bound is expected to get
reduced by several orders of magnitude, in which case, it could signal a new and
intermediate length scale between the electroweak and the Planck scale.

\section{The Landau Levels}

Next consider a particle of mass $m$ and charge $e$ in a constant
magnetic field ${\vec B} = B {\hat z}$, described by the vector
potential ${\vec A}=Bx {\hat y}$ in the Landau gauge.
The corresponding Hamiltonian is
\bea H_0 &=& \frac{1}{2m}\le( \vec p - e \vec A\ri)^2  \la{lanham1}\\
&=& \frac{p_x^2}{2m} + \frac{p_y^2}{2m} - \frac{eB}{m}~x p_y +
\frac{e^2 B^2}{2m}~x^2~. \la{lanham2}
\eea
Since $p_y$ commutes with $H$, replacing it with its eigenvalue
$\hbar k$, we get
\be H_0 = \frac{p_x^2}{2m} + \frac{1}{2} m \omega_c^2 \le( x -
\frac{\hbar k}{m \omega_c}\ri)^2 ~,\la{lanham4}\ee
where $\omega_c=eB/m$ is the cyclotron frequency. This is nothing
but the Hamiltonian of a harmonic oscillator in the $x$ direction,
with its equilibrium position given by $x_0 \equiv \hbar k/m
\omega_c$. Consequently, the eigenfunctions and eigenvalues are
given by
\bea \psi_{k,n} (x,y) &=& e^{ik y} \phi_n (x-x_0) \\
E_n &=& \hbar \omega_c \le( n +\frac{1}{2} \ri)~,~n\in N~, \eea
where $\phi_n$ are the harmonic oscillator wave-functions.

Following the procedure outlined in Appendix A, the
GUP corrected Lagrangian, coupled minimally to a $U(1)$ gauge
potential yields the GUP corrected Hamiltonian after a Legendre
transformation. The final result is [Eq.(\ref{gupmincoup})]
\bea H &=& \frac{1}{2m}\le( \vec p - e \vec A\ri)^2 +
\frac{\beta}{m}\le( \vec p - e \vec A\ri)^4 \nn \\
%
&=& H_0 + 4\b m H_0^2
%
\eea
where in the last step we have inverted Eq.(\ref{lanham1}) to
write $(\vec p - e\vec A)$ in terms of $H_0$.
Evidently, the eigenfunctions remain unchanged. This alone guarantees, for
example, that the GUP will have no effect at all on phenomena
such as the Quantum Hall Effect \cite{laughlin}, the Bohm-Aharonov effect \cite{Aharonov:1959fk},
and Dirac Quantization \cite{dirac}.
However, the eigenvalues shift by
\bea
\Delta E_{n(GUP)} &=& 4\b m \langle \phi_n |H_0^2| \phi_n \rangle \nn\\
&=& 4\b m (\hbar \omega_c )^2 \le( n + \frac{1}{2} \ri)^2 ~,\\
\mbox{or}~~
\frac{\Delta E_{n(GUP)}}{E_n} &=& 4\b m (\hbar \omega_c ) \le( n + \frac{1}{2} \ri) \\
&\approx& \beta_0 \frac{m}{M_{Pl}} \frac{\hbar \omega_c}{M_{Pl}c^2}~.
\eea
%
%

For an electron in a magnetic field of $10~T$, $\omega_c \approx 10^3~GHz$ and we get
%
%
\bea
\frac{\Delta E_{n(GUP)}}{E_n}
&\approx&  \le( 0.42 \times 10^{-22} \ri) \times \le(5.48\times 10^{-32} \ri)\b_0 \nn\\
&=& 2.30 \times 10^{-54} \b_0~.
\eea
Thus, Quantum Gravity/GUP does affect the Landau levels. However,
once again, assuming $\beta_0 \sim 1$ renders the correction too small
to be measured. Without this assumption,
an accuracy of $1$ part in $10^3$ in direct measurements
of Landau levels using a Scanning Tunnel Microscope (STM) (which is somewhat
optimistic) \cite{wildoer}, the upper bound on $\b_0$ follows
\be
\b_0 < 10^{50}~.
\la{beta2}
\ee
This bound is far weaker than that set by electroweak measurements, but
compatible with the latter (as was the case for the Lamb shift).
Once again, it is expected that the above accuracy will increase significantly with time,
predicting a tighter bound on $\b_0$, as well as perhaps an intermediate length
scale.



\section{Potential Step}

Next, we study the one dimensional potential step given by
\be
V(x) = V_0~\theta(x)~,
\ee
where $\theta(x)$ is the usual step function.
Assuming $E>V_0$, the Schr\"odinger equation to the left and right of the barrier
are given respectively by
\bea
&& d^2\psi_< + k^2 \psi_< - \ell_{Pl}^2 d^4 \psi_< = 0~~,~x \leq 0 \\
&& d^2\psi_> + k_1^2 \psi_> - \ell_{Pl}^2 d^4 \psi_> = 0~~,~ x \geq 0 \\
k&=&\sqrt{2mE/\hbar^2}~,k_1=\sqrt{2m(E-V_0)/\hbar^2}~,
\eea
where $d^n \equiv d^n/dx^n$. Assuming solutions of the form
$\psi_{<,>}=e^{mx}$, we get
\bea
&& m^2 + k^2  - \ell_{Pl}^2 m^4 = 0~~,~x \leq 0 \\
&& m^2 + k_1^2  - \ell_{Pl}^2 m^4 = 0~~,~x \geq 0
\eea
with the following solution sets to leading order in
$\b$, each consisting of $4$ values of $m$
\bea
x \leq 0: m\hspace{-0.5ex}&=&\hspace{-1ex}\{ \pm ik', \pm 1/\ell_{Pl} \}~,k'\equiv k(1-\beta \hbar^2 k^2 ) \la{kprimed1} \\
x \geq 0: m\hspace{-0.5ex}&=&\hspace{-1ex}\{ \pm ik_1', \pm 1/\ell_{Pl} \}~,k_1'\equiv k_1(1-\beta \hbar^2 k_1^2 ) \la{k1primed1}
\eea
and the wavefunctions
\bea
\psi_< &=& Ae^{ik'x} + B e^{-ik'x} + A_1 e^{x/\ell_{Pl}}~~,~~x \leq 0 \la{steppsi1} \\
\psi_> &=& Ce^{ik_1'x} + D_1 e^{-x/\ell_{Pl}}~~,~~x\geq 0 \la{steppsi2}
\eea
where we have omitted the left-mover from $\psi_>$ and the
exponentially growing terms from both $\psi_<$ and $\psi_>$. Note
that the $\ell_{Pl}$-dependent decaying terms are a result of the GUP
induced fourth order Schr\"odinger equation. They are independent
of both $E$ and $V_0$, and appear to be {\it non-perturbative} in
nature. Now the boundary conditions at $x=0$ consist of $4$
equations (instead of the usual $2$)
\be
d^n\psi_<|_0 = d^n\psi_>|_0~~,~n=0,1,2,3~,
\ee
giving rise to the following
\bea
A + B + A_1 &=& C + D_1 \\
ik'(A-B) + \frac{A_1}{\ell_{Pl}} &=& ik_1'C - \frac{D_1}{\ell_{Pl}} \\
-k'^2(A+B) + \frac{A_1}{\ell_{Pl}^2} &=& -k_1'^2 C + \frac{D_1}{\ell_{Pl}^2} \\
-ik'^3 (A-B) + \frac{A_1}{\ell_{Pl}^3} &=& - ik_1'^3 C - \frac{D_1}{\ell_{Pl}^3}~.
\eea
The above equations have the following solutions to leading order in $\beta$
\bea
\frac{B}{A} &=& \frac{k'-k_1'}{k'+k_1'} \\
\frac{C}{A} &=& \frac{2k'}{k'+k_1'} \\
\frac{A_1}{A} &=& k'(k'-k_1')\ell_{Pl}^2 \\
\frac{D_1}{A} &=& -k'(k'-k_1')\ell_{Pl}^2~.
\eea
Note that $A_1$ and $D_1$ are of the order $\ell_{Pl}^2$ ($\sim \beta$), and that they vanish for
$V_0=0$ (when $k'=k_1'$). In other words, the decaying terms are absent for the free
particle. Computing the conserved current using (\ref{newcurrent1}), we get
\bea
J_< &=& k'\le[ |A|^2 - |B|^2 \ri] \\
J_> &=& k_1' |C|^2~.
\eea
Naturally, the reflection and transmission coefficients are defined as
\bea
R &=& \le| \frac{B}{A} \ri|^2 =  \le( \frac{k'-k_1'}{k'+k_1'} \ri)^2 \\
&=&  \le( \frac{k-k_1}{k+k_1} \ri)^2 \le( 1- 4\b \hbar^2 k k_1\ri) \la{reflex2}
\\
T &=& \frac{k_1'}{k'} \le| \frac{C}{A} \ri|^2 = \le( \frac{2k'}{k'+k_1'} \ri)^2 \\
&=& \frac{4 kk_1}{(k+k_1)^2} \le( 1 + \beta \hbar^2 (k - k_1 )^2 \ri) \la{trans2}
\\
R + T &=& 1~, \la{rt1}
\eea
Note that the GUP affects both $R$ and $T$. In deriving Eqs.(\ref{reflex2}) and (\ref{trans2}),
we have used Eqs.(\ref{kprimed1}) and (\ref{k1primed1}) to leading order in $\b$. Also,
the conservation equation (\ref{rt1}) would not hold if we had not included the exponential solutions
in Eqs.(\ref{steppsi1}-\ref{steppsi2}).


\section{Potential barrier}

A potential barrier of height $V_0$ from $x=0$ and
$x=a$ in Eq.(\ref{gupham1}) is given by
\be
V(x) = V_0~\le[ \theta(x) - \theta(x-a) \ri]~,
\ee
where $\theta(x)$ is the usual step function.
In this case, we assume $E<V_0$.
The Schr\"odinger equation in the three regions
(which, henceforth, are denoted for brevity  $R_1$, $R_2$, and $R_3$
for $x \leq 0$,  $0\leq x \leq a$, and  $x \geq a$, respectively,)
are given respectively by
\bea
&& d^2\psi_< + k^2 \psi_< - \ell_{Pl}^2 d^4 \psi_< = 0~,~\mbox{in} ~R_1 \\
&& d^2\psi_> + k_1^2 \psi_> - \ell_{Pl}^2 d^4 \psi_> = 0~,~\mbox{in} ~R_2 \\
&& d^2\psi_{>>} + k^2 \psi_{>>} - \ell_{Pl}^2 d^4 \psi_{>>} = 0~,~\mbox{in} ~R_3 \\
k&=&\sqrt{2mE/\hbar^2}~,~k_1=\sqrt{2m(V_0-E)/\hbar^2}~, 
\eea
where $d^n \equiv d^n/dx^n$. Assuming solutions of the form
$\psi_{<,>,>>}=e^{mx}$, we get
\bea
&& m^2 + k^2  - \ell_{Pl}^2 m^4 = 0~~,~\mbox{in} ~R_1 \\
&& m^2 + k_1^2  - \ell_{Pl}^2 m^4 = 0~~,~\mbox{in} ~R_2 \\
&& m^2 + k^2  - \ell_{Pl}^2 m^4 = 0~~,~\mbox{in} ~R_3
\eea
with the following solution sets to leading order in
$\b$ and $\ell_{Pl}$, each consisting of $4$ values of $m$
\bea
R_{1}: m &=& \{ \pm ik', \pm 1/\ell_{Pl} \}~,k'\equiv k(1-\beta \hbar^2 k^2 )~, \\
R_{2}: m &=& \{ \pm ik_1', \pm 1/\ell_{Pl} \}~,k_1'\equiv k_1(1-\beta \hbar^2 k_1^2 )~, \\
R_{3}: m &=& \{ \pm ik', \pm 1/\ell_{Pl} \}~,k'\equiv k(1-\beta \hbar^2 k^2 )~,
\eea
and the wavefunctions in $R_1$, $R_2$, and $R_3$, respectively, are
\bea
\psi_< &=& Ae^{ik'x} + B e^{-ik'x} + A_1 e^{x/\ell_{Pl}}~, \la{barrierpsi1} \\
\psi_> &=& Fe^{k_1'x} + Ge^{-k_1'x} + H_1e^{x/\ell_{Pl}} + L_1 e^{-x/\ell_{Pl}}~\hspace{-1.5ex}, \la{barrierpsi2} \\
\psi_{>>} &=& Ce^{ik_1'x} + D_1 e^{-x/\ell_{Pl}}~,  \la{barrierpsi3}
\eea
where we have omitted the left-mover from $\psi_{>>}$ and the exponentially growing
terms from both $\psi_<$ and $\psi_{>>}$.
Note once again the $\ell_{Pl}$-dependent decaying terms.
Now the boundary conditions consist of $8$ equations, $4$ each from
$x=0$ and $x=a$
\bea
d^n\psi_<|_0 &=& d^n\psi_>|_0~~,~n=0,1,2,3~, \\
d^n\psi_>|_a &=& d^n\psi_{>>}|_a~~,~n=0,1,2,3~,
\eea
giving rise to the following
\bea
A + B + A_1 &=& F + G + H + L_1 \\
ik'(A-B) + \frac{A_1}{\ell_{Pl}} &=& k_1'(F-G) + \frac{H_1-L_1}{\ell_{Pl}} \\
-k'^2(A+B) + \frac{A_1}{\ell_{Pl}^2} &=& k_1'^2 (F+G) + \frac{H_1+L_1}{\ell_{Pl}^2} \\
-ik'^3 (A-B) + \frac{A_1}{\ell_{Pl}^3} &=& k_1'^3 (F-G) + \frac{H_1-L_1}{\ell_{Pl}^3}~ \\
F e^{k_1'a} + G e^{-k_1'a} &+& H_1 e^{a/\ell_{Pl}} + L_1e^{-a/\ell_{Pl}} \nn\\
&=& C e^{ik'a} + D_1 e^{-a/\ell_{Pl}} \\
k_1'\le(F e^{k_1'a} - G e^{-k_1'a}\ri) &+& \frac{H_1}{\ell_{Pl}} e^{a/\ell_{Pl}} - \frac{L_1}{\ell_{Pl}}e^{-a/\ell_{Pl}} \nn\\
&=& ik' C e^{ik'a} - \frac{D_1}{\ell_{Pl}} e^{-a/\ell_{Pl}} \\
k_1'^2\le(F e^{k_1'a} + G e^{-k_1'a}\ri) &+& \frac{H_1}{\ell_{Pl}^2} e^{a/\ell_{Pl}} + \frac{L_1}{\ell_{Pl}^2}e^{-a/\ell_{Pl}} \nn\\
&=&\hspace{-1ex}-k'^2 C e^{ik'a} +\hspace{-0.5ex} \frac{D_1}{\ell_{Pl}^2} e^{-a/\ell_{Pl}} \\
k_1'^3\le(F e^{k_1'a} - G e^{-k_1'a}\ri) &+& \frac{H_1}{\ell_{Pl}^3} e^{a/\ell_{Pl}} - \frac{L_1}{\ell_{Pl}^3}e^{-a/\ell_{Pl}} \nn\\
&=&\hspace{-1ex}-ik'^3 C e^{ik'a}\hspace{-1ex} - \hspace{-0.5ex}\frac{D_1}{\ell_{Pl}^3} e^{-a/\ell_{Pl}}\hspace{-1ex}.
\eea
These have the solutions to leading order in $\beta$
\bea
\frac{B}{A} &=& \frac{ (k'^2+k_1'^2)(e^{2k_1'a} -1) }{e^{2k_1'a}(k'+ik_1')^2 - (k'-ik_1')^2}
\\
\frac{C}{A} &=& \frac{ 4i k'k_1'e^{-ik'a} e^{k_1'a} }{e^{2k_1'a}(k'+ik_1')^2 - (k'-ik_1')^2}
\\
\frac{F}{A} &=& \frac{ -2k'(k'-ik_1') }{e^{2k_1'a}(k'+ik_1')^2 - (k'-ik_1')^2}
\\
\frac{G}{A} &=& \frac{ 2 e^{2k_1'a} k' (k'+ ik_1') }{e^{2k_1'a}(k'+ik_1')^2 - (k'-ik_1')^2}~.
\eea
\par\noindent
Computing the conserved current as given in (\ref{newcurrent1}), we get
\bea
J_< &=& k'\le[ |A|^2 - |B|^2 \ri] \\
J_{>>} &=& k' |C|^2~.
\eea
Thus, the reflection and transmission coefficients are given by
\bea
R &=& \le| \frac{B}{A} \ri|^2 \hspace{-1ex}=
\le| \frac{ (k'^2+k_1'^2)(e^{2k_1'a} -1) }{e^{2k_1'a}(k'+ik_1')^2 - (k'-ik_1')^2} \ri|^2
\\
&=&
\le[ 1 + \frac{(2k'k_1')^2}{(k'^2+k_1'^2)^2\sinh^2(k_1'a)} \ri]^{-1}
\la{reflex3}
\\
T &=& \le| \frac{C}{A} \ri|^2 \hspace{-1ex}=
\le| \frac{ 4i k'k_1'e^{-ik'a} e^{k_1'a} }{e^{2k_1'a}(k'+ik_1')^2 - (k'-ik_1')^2} \ri|^2
\\
&=&
\le[
1 + \frac{(k'^2 + k_1'^2)^2\sinh^2(k_1'a)}{(2k'k_1')^2}
\ri]^{-1}
\la{trans3}
\\
R + T &=& 1~. \la{rt2}
\eea
Note that the GUP affects both $R$ and $T$. Once again,
the conservation equation (\ref{rt2}) would not hold if we had not included the exponential solutions
in Eqs(\ref{barrierpsi1}-\ref{barrierpsi3}).

{}From Eq.(\ref{trans3}) above, and using the definitions of $k,k_1,k',k_1'$,
it can be shown that when $k_1 a \gg 1$, the transmission coefficient
is approximately
\bea
T&=&T_0 \le[
1 + \frac{4m\b (2E-V_0)^2}{V_0} \ri.\nn\\
&&+\le. \frac{2\b a}{\hbar} [2m (V_0-E)]^{{3}/{2}} \ri]~,\la{stmt} \\
T_0 &=&
\frac{16E(V_0-E)}{V_0^2} e^{-2k_1a}~,
\eea
$T_0$ being the `usual' tunnelling amplitude. Now $T$ is proportional to the
current $I$ flowing between the tip and a sample in a Scanning Tunnel Microscope (STM).
The current is usually amplified using an amplifier of gain ${\cal G}$.
Thus, the enhancement in current due to GUP is given by
\bea
\frac{\d I}{I_0} &=& \frac{\d T}{T_0}
= 4\b_0 \frac{m (2E-V_0)^2}{V_0} + \frac{2\b a}{\hbar} [2m (V_0-E)]^{{3}/{2}} \nn\\
&=& \frac{4\b_0 m}{M_{Pl}} \frac{(2E-V_0)^2/V_0}{E_{Pl}} \nn \\
&+& 4\sqrt{2} \b_0 \frac{a}{\ell_{Pl}}
\left( \frac{m}{M_{Pl}} \ri)^{3/2}
\left( \frac{V_0-E}{E_{Pl}} \ri)^{3/2}~.
\eea

\par\noindent
Then, assuming the following approximate (but realistic) values \cite{stroscio}
\bea
m &=& m_e = 0.5~MeV/c^2 \\
E, V_0 &=& 10~eV  \\ 
a &=& 10^{-10} ~m \\
I &=& 10^{-9} ~A \\
{\cal G} &=& 10^9
\eea
we get
\bea
k_1 &=& 10^{10}~m^{-1} \\
T_0 &=&  e^{-1}     \\
\frac{\d I}{I_0} &=& \frac{\d T}{T_0} = 10^{-48} \b_0  \\
\d {\cal I} &\equiv& {\cal G} \d I = 10^{-48} \b_0~A ~.
\eea
Thus, due to the excess current $\d{\cal I}$ added up to the charge of just one electron,
$e=10^{-19}~C$, one would have to wait for a time
\be
\tau = \frac{e}{\d{\cal I}}  = 10^{29} \b_0^{-1}~s~.
\ee
If $\b_0 \approx 1$, this is much bigger than the age of our universe ($10^{18}~s$)!
However, if the quantity $\d {\cal I}$ can be increased
by a factor of about $10^{21}$, say by a combination of increase in
$I$ and ${\cal G}$, and by a larger value of $\b_0$, the above time will be reduced
to about a year ($\approx 10^8~s$), and one can hope that the effect of GUP can be measured.

Conversely, if such a GUP induced current cannot be
measured in such a time-scale, it will put an upper bound
\be
\b_0 < 10^{21}~. \la{beta3}
\ee
Note that this is more stringent than the two previous examples, and
is in fact consistent with that set by the electroweak scale!
In practice however, it may be easier to experimentally determine the
{\it apparent barrier height} $\Phi_A \equiv V_0-E$,
and the (logarithmic) rate of increase of current with the gap. From Eq.(\ref{stmt})
they are related by \cite{stroscio}
\be
\sqrt{\Phi_A} = \frac{\hbar}{\sqrt{8m}}
\le|\frac{d\ln I}{da}\ri| \le( 1 - \frac{\beta \hbar^2}{4}
\le| \frac{d\ln I}{da}\ri|^2 \ri)~.
\ee
\par\noindent
The cubic deviation from the linear $\sqrt{\Phi_A}$ vs
$\le| \frac{d\ln I}{da} \ri|$ curve predicted by GUP may be easier to spot
and the value of $\b$ estimated with improved accuracies.


\section{Discussion}
The above analysis, especially Eqs.(\ref{beta1}), (\ref{beta2})
and (\ref{beta3}) indicate that
a much larger coefficient of the additional term in the GUP (than previously
thought) is not ruled out by current observations. These translate to
intermediate length scales
$\ell_{inter} \sim 10^{18} \ell_{Pl},
10^{25} \ell_{Pl}~\mbox{and}~10^{10} \ell_{Pl}
$
respectively, of which the first two are far bigger than the electroweak scale,
and the last, although smaller, may get further constrained with increased accuracies.
In any case, more accurate measurements of the quantum phenomena studied here, or others,
are required to tighten the above bounds. Then one might be able to see whether a
true intermediate length scale emerges.
It is not inconceivable that such a new length scale may
show up in future experiments in the LHC.
On the other hand, it is quite possible that $\beta_0 \sim 1$, the
effects of GUP on low energy phenomena are negligible, and there is no
intermediate length scale, supporting a recent argument \cite{Shaposhnikov:2007nj}.

Perhaps more importantly, our study reveals the universality of GUP effects,
meaning that the latter can potentially be tested in a wide class of quantum mechanical
systems, in which they maybe more pronounced. Possibilities include statistical
mechanical systems (where a large number of particles may help in the enhancement),
study of normally forbidden quantum processes to see if the GUP allows them,
systems which may be affected by a fractional power of $\beta$, and GUP effects in
cosmology. Any signature of testable predictions in one or more of the above
(or perhaps others) could open a much needed low-energy
`window' to Quantum Gravity Phenomenology.


\no {\bf Acknowledgment}

We thank K. Ali, B. Belchev, A.
Dasgupta, R. B. Mann, S. Sur and M. Walton for useful
discussions, and the anonymous referees for
useful comments. This work was supported in part by the Natural
Sciences and Engineering Research Council of Canada and by the
Perimeter Institute for Theoretical Physics.


\hs{-1cm}
\appendix{\bf {Appendix A: Minimal Coupling to Electromagnetism}}

For the Hamiltonian (\ref{gupham1}) the equations of motion are
\bea
\dot x_0 = \frac{\pa H}{\pa p} = \frac{p_0}{m} + \frac{4\beta}{m} {p_0}^3 \la{hameqn1}
\eea
{its inverse~}
\bea
p_0 &=& m \dot x_0 - 4\beta (m \dot x_0)^3 \la{hameqn1a} \\
\mbox{and}~~\dot p_0 &=& - \frac{\pa H}{\pa x_0} = -\frac{\pa V}{\pa x_0} \la{hameqn2}~.
\eea
From Eqs.(\ref{hameqn1}) and (\ref{hameqn1a}), we get the modified Newton's law
\be
m\ddot x_0 \le( 1 -12 \beta m^2 \dot x_0^2  \ri) = -\frac{\pa V}{\pa x_0}~,
\ee
which is also the Euler-Lagrangian equation for the following GUP modified Lagrangian
\be
L = p_0 \dot x_0 - H = \frac{1}{2} m \dot x_0^2 - V(x_0) - \frac{\b}{m} (m \dot x_0)^{4}
~.
\ee

Following the procedure outlined in \cite{LLCTF}, Section 16, we couple the
above non-relativistic one-dimensional
Lagrangian to a $U(1)$ gauge potential $A^\mu=(\phi, A)$
(relativistic and higher dimensional generalizations are straightforward)
%
%
\be
L = \frac{1}{2} m \dot x_0^2 - V(x) - \frac{\b}{m} (m \dot x_0)^4
+ e \le( A \dot x_0 - \phi \ri)~.
\ee
The generalizations of (\ref{hameqn1a}) and (\ref{gupham1}) are
\bea
p_0 &=& m\dot x_0 + e A - 4\b m^3 \dot x_0^3 \\
H &=& \frac{1}{2} m \dot x_0^2 +e\phi - 3\b m^3 \dot x_0^4~.
\eea
Eliminating $\dot x_0$ in favor of $p$, it follows that
\bea
H &=& e\phi + \frac{1}{2m}\le( p_0 - e A \ri)^2 + \frac{\b}{m} \le( p_0 - e A\ri)^4
\label{gupmincoup}\\
&\equiv& H_0 + H_{1} ~.
\eea

Note that for $\phi=0$, the eigenfunctions of $H_0$ and $H$ are identical, albeit with
different eigenvalues. We have used this fact in the section on Landau levels. Also,
the above form ensures that the
classical gauge invariance of the action
translates into the multiplication by a
phase of the quantum wavefunction, under a gauge transformation.





\begin{thebibliography}{99}


\bibitem{qgpheno}
See e.g. G. Amelino-Camelia, arXiv:0806.0339; D. Sudarsky, Int. J.
Mod. Phys. {\bf D14} (2005) 2069-2094 [arXiv:gr-qc/0512013]; D.
Kimberly, J. Magueijo, arXiv:gr-qc/0502110; M. Bojowald, H. A.
Morales-T\'ecotl, H. Sahlmann, Phys. Rev. {\bf D71} (2005) 084012
[arXiv:gr-qc/0411001]; Y. Jack-Ng, arXiv:gr-qc/0405078;
T.~Jacobson, S.~Liberati and D.~Mattingly, Springer Proc.\ Phys.\  {\bf 98}, 83 (2005) [arXiv:gr-qc/0404067];
F. Dowker, J. Henson, R. D. Sorkin, Mod. Phys. Lett. {\bf A19} (2004) 1829-1840 [arXiv:gr-qc/0311055];
M. Chaichian, M. M. Sheikh-Jabbari, A. Tureanu, Phys. Rev. Lett. {\bf 86} (2001) 2716 [arXiv:hep-th/0010175];
F.~Scardigli, Phys.\ Lett.\  B {\bf 452}, 39 (1999)[arXiv:hep-th/9904025].

\bibitem{sdecvprl}
S.~Das and E.~C.~Vagenas,
  Phys.\ Rev.\ Lett.\  {\bf 101}, 221301 (2008)
  [arXiv:0810.5333 [hep-th]].


\bibitem{guppapers} D. Amati, M. Ciafaloni, G. Veneziano,
Phys. Lett. {\bf B 216} (1989) 41;
M.~Maggiore,
  Phys.\ Lett.\  B {\bf 304} (1993) 65  [arXiv:hep-th/9301067];
M.~Maggiore,
  Phys.\ Rev.\  D {\bf 49} (1994) 5182
  [arXiv:hep-th/9305163];
M.~Maggiore,
  Phys.\ Lett.\  B {\bf 319} (1993) 83
  [arXiv:hep-th/9309034];
L.~J.~Garay, Int.\ J.\ Mod.\ Phys.\  A {\bf 10} (1995) 145 [arXiv:gr-qc/9403008];
S.~Hossenfelder, M.~Bleicher, S.~Hofmann, J.~Ruppert, S.~Scherer and H.~Stoecker,
    Phys.\ Lett.\  B {\bf 575}, 85 (2003)
  [arXiv:hep-th/0305262].


\bibitem{Harbach:2003qz}
  U.~Harbach, S.~Hossenfelder, M.~Bleicher and H.~Stoecker, Phys.\ Lett.\  B {\bf 584}, 109 (2004)  [arXiv:hep-ph/0308138];
  S.~Hossenfelder,  Mod.\ Phys.\ Lett.\  A {\bf 19}, 2727 (2004)  [arXiv:hep-ph/0410122];
  U.~Harbach and S.~Hossenfelder,  Phys.\ Lett.\  B {\bf 632}, 379 (2006)  [arXiv:hep-th/0502142].




\bibitem{kmm} A. Kempf, G. Mangano, R. B. Mann, Phys. Rev. {\bf
D52} (1995) 1108 [arXiv:hep-th/9412167]; S.~Benczik, L.~N.~Chang,
D.~Minic and T.~Takeuchi,
  Phys.\ Rev.\  A {\bf 72}, 012104 (2005)
  [arXiv:hep-th/0502222].


\bibitem{kempf} A. Kempf, J.Phys. {\bf A 30} (1997) 2093 [arXiv:hep-th/9604045].



\bibitem{brau}
 F.~Brau,
  J.\ Phys.\ A  {\bf 32}, 7691 (1999)
  [arXiv:quant-ph/9905033].



%
%

\bibitem{bransden} B. H. Bransden, C. J. Joachain, {\it Quantum
Mechanics}, Second Edition, Pearson Education, Delhi (2007).

\bibitem{bd} J. D. Bjorken, S. D. Drell, {\it Relativistic Quantum
Mechanics}, Mc-Graw Hill, New York, (1964) p.60.



\bibitem{newton} G. Newton, D. A. Andrews, P. J. Unsworth,
Phil. Trans. Roy. Soc. Lond. Series A,
Math. and Phys. Sc., {\bf 290}, No. 1373 (1979), 373.

\bibitem{laughlin} R. B. Laughlin, Phys. Rev. {\bf B 23} (1981) 5632.

\bibitem{Aharonov:1959fk}
  Y.~Aharonov and D.~Bohm,
  Phys.\ Rev.\  {\bf 115} (1959) 485;
  Y.~Aharonov and D.~Bohm,
  Phys.\ Rev.\  {\bf 123} (1961) 1511.

\bibitem{dirac}
P. A. M. Dirac, Proc. Roy. Soc. {\bf A 133} (1931) 610.


\bibitem{wildoer} J. W. G. Wild\"oer, C. J. P. M. Harmans,
H. van Kempen, Phys. Rev. {\bf B 55} (1997) R16013.


\bibitem{stroscio}
J. A. Stroscio, W. J. Kaiser,
{\it Scanning Tunneling Microscopy}, Academic Press, 1993.


\bibitem{Shaposhnikov:2007nj}
  M.~Shaposhnikov,
  {\it ``Is there a new physics between electroweak and Planck scales?''},
  arXiv:0708.3550 [hep-th].


\bibitem{LLCTF} L. D. Landau and E. M. Lifshitz, {\it The Classical Theory of
Fields}, 4th Revised English Edition, Pergamon (1989),
Oxford.



\end{thebibliography}
\end{document}